\documentclass[runningheads]{llncs}
\usepackage{graphicx}
\usepackage{booktabs}
\usepackage{nicefrac}
\usepackage{balance}
\usepackage{color}
\usepackage{subfig}
\PassOptionsToPackage{hyphens}{url}
\usepackage{hyperref}
\begin{document}
\title{Impact of Online Health Awareness Campaign: \\Case of National Eating Disorders Association}
\titlerunning{Impact of Online Health Awareness Campaign}
%
\author{Yelena Mejova \inst{1}\orcidID{0000-0001-5560-4109} \and\\
V\'{i}ctor Suarez-Lled\'{o}\inst{2}\orcidID{0000-0001-7714-6719}}
\authorrunning{Mejova \& Suarez-Lled\'{o}}
%
\institute{ISI Foundation, Turin, Italy \and
University of Cadiz, Cadiz, Spain\\
\email{yelenamejova@acm.org}, \email{victor.sanz@uca.es}\vspace{-0.5cm}}

\maketitle              
\begin{abstract}
National Eating Disorders Association conducts a NEDAwareness week every year, during which it publishes content on social media and news aimed to raise awareness of eating disorders. Measuring the impact of these actions is vital for maximizing the effectiveness of such interventions. This paper is an effort to model the change in behavior of users who engage with NEDAwareness content. We find that, despite popular influencers being involved in the campaign, it is governmental and nonprofit accounts that attract the most retweets. Furthermore, examining the tweeting language of users engaged with this content, we find linguistic categories concerning women, family, and anxiety to be mentioned more within the 15 days after the intervention, and categories concerning affiliation, references to others, and positive emotion mentioned less. We conclude with actionable implications for future campaigns and discussion of the method's limitations.

\end{abstract}

\section{Introduction}

The National Eating Disorders Association (NEDA) conducts a ``NEDAwareness'' week every year at the end of February\footnote{\url{https://www.nationaleatingdisorders.org/blog/announcing-national-eating-disorders-awareness-week-2020}}. NEDA is a nonprofit organization dedicated to supporting individuals and families affected by eating disorders, focusing on prevention, cures and access to quality care. Millions of people in US are at some point affected by eating disorders, which have a second highest mortality rate for mental disorders \cite{Levine2003a}. During NEDAwareness week, NEDA publishes content on social media and news, promoting awareness and linking to resources. 

As health intervention campaigns on social media are becoming more prevalent \cite{Chou2018}, the evaluation of their impact is becoming imperative in conducting effective interventions. This project is an effort to understand the impact NEDAwareness content has on social media users, in particular those using Twitter. We collect a dataset of NEDA-related tweets during the 2019 NEDAwareness week, capturing 20,197 tweets from 11,470 users. We extend the dataset with historical tweets of 7,870 users in order to capture their behavior before and after the event, and complement this collection with a baseline control group of 1,668 users.
We begin by analyzing the dissemination and scope of this campaign, finding that, despite having influential accounts involved, it is the government agencies and nonprofits that achieve the most retweets. Secondly, we attempt to capture to what extent NEDA content achieves significant changes in the conversation topics of its audience. Compared to baseline users who did not engage with NEDA content, we find the users who did engage significantly change their language in the two weeks after the intervention, focusing more on womanhood, family, and anxiety, and sharing fewer social experiences and positive emotions. 

\section{Related Work}
\vspace{-0.1cm}

Social media has been widely employed to study mental health, ranging from fitness enthusiasts \cite{Holland2017a} and pro-eating disorder communities \cite{chancellor2016quantifying} to depression \cite{de2013predicting} and suicidal ideation \cite{de2016discovering}. Language of social media posts has been successfully used to predict their authors' internal states, achieving, for instance, precision of 0.74 on CES-D depression scale questionnaire \cite{de2013predicting,radloff1977ces}. Images have also been used to characterize those with depression and anxiety \cite{guntuku2019twitter}.

Meanwhile, health authorities began employing social media as a platform for behavior change and health awareness campaigns, especially in the domain of cancer awareness. Yearly drives \#WorldCancerDay and National Breast Cancer Awareness Month (NBCAM) have been shown to attract engagement especially around women's cancers, and especially on Twitter \cite{Vraga2018b}. Cancer discussions have been analyzed in the form of a follow network, showing distinct communities of breast and prostate cancer conversations \cite{Himelboim2014a}. Communities around pro- and anti- eating disorder have also been examined on Flickr \cite{yom2012pro}, Tumblr \cite{de2015anorexia} and Instagram \cite{chancellor2016quantifying}. Most of these works, however, measure the interaction of the audience with the campaign material, failing to follow up on the potential changes in behavior after the intervention. A notable exception is the measurement of whether those posting to pro-anorexia Flickr communities continue to do so after being exposed to anti-anorexia content \cite{yom2012pro}. They find that, unlike the intended effect, these users would post longer to pro-anorexia communities. Thus, it is important to verify the effects of online interventions. In this study, we attempt to quantify the change in posting activity of those engaged in NEDAwareness week in the time following the intervention. 

Beyond campaign engagement, recent causality methods have encouraged the measurement of post-intervention behavior. A general framework for event outcome analytics through social media has been proposed by \cite{olteanu2017distilling,kiciman2018answering}, and has been employed to, for instance, gauge the impact of alcohol use in students \cite{kiciman2018using} and psychopathological effects of psychiatric medication \cite{saha2019social}. Additionally, West et al. \cite{west2013cookies} used search queries to track the behavior of users after signaling weight loss intention. In this work, we employ temporal analysis of daily online interactions, compared to a non-treatment control, to uncover outcomes of a communication campaign.


\section{NEDAwareness Week Data Collection}

\begin{figure}[t]
    \centering
    \includegraphics[width=0.7\linewidth]{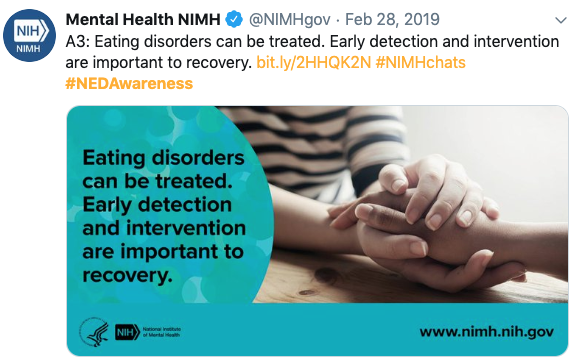}
    \caption{Example post during NEDAwareness week.\vspace{-0.3cm}}
    \label{fig:nedaweekpost}
\end{figure}


Primary data collection happened during the intervention week February 25 - March 3 using Twitter Streaming API with the following hashtags: \texttt{NEDAstaff}, \texttt{NEDAwareness}, \texttt{NEDA}, \texttt{ComeAsYouAre}, \texttt{SOSChat} (compiled with the assistance of NEDA staff). The resulting collection comprises of 20,197 tweets from 11,470 users. We then performed another collection of the historical tweets on April 18-21, 2019, resulting in up to 3,200 tweets for 7,870 of the users (some accounts were closed or private). Thus, we obtained approximately a total of 12 million tweets from all users. 

\begin{figure}[t]
    \centering
    \includegraphics[width=0.8\linewidth]{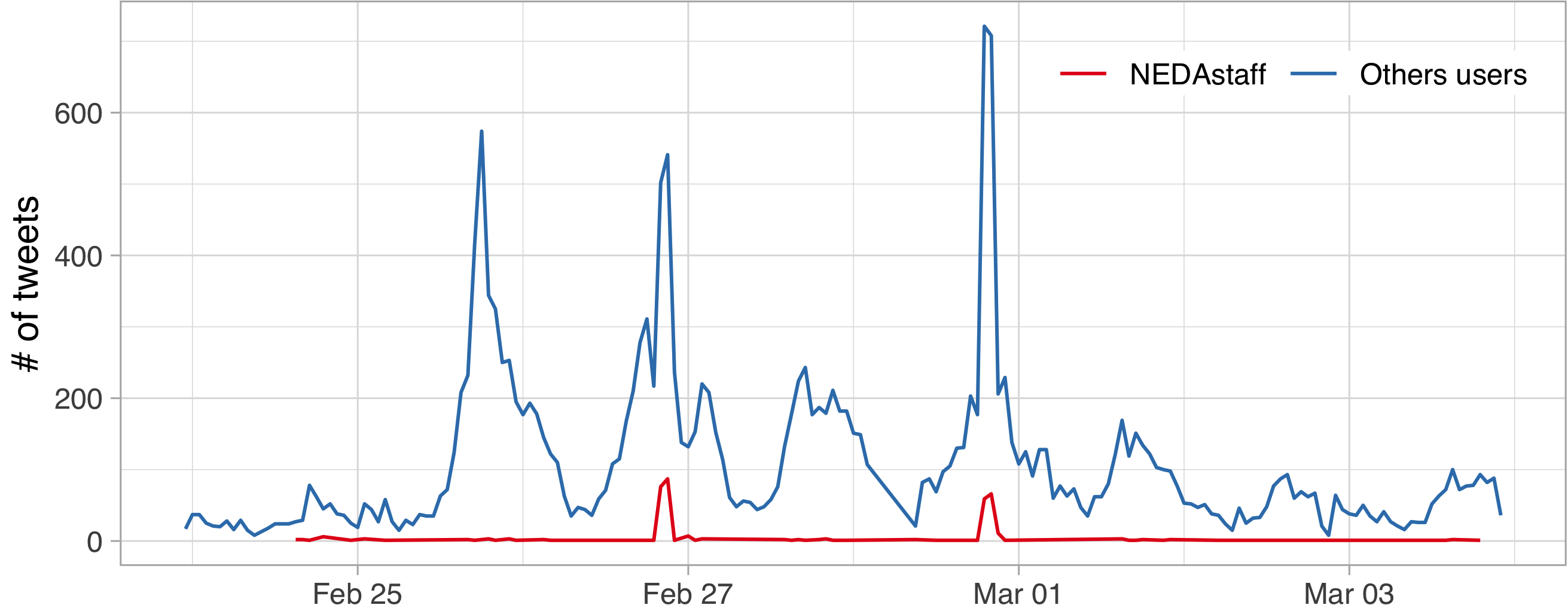}
    \caption{Tweets per hour during the NEDAwareness week.}
    \label{fig:nedaweek}
\end{figure}

An example tweet is shown in Figure \ref{fig:nedaweekpost}. As the theme of 2019 was ``Come as you are'', the content often deals with mental health issues of people from a plurality of ages, races, genders and gender identities. As can be seen in Figure \ref{fig:nedaweek}, the campaign spans several weeks, with much of the content being produced not by the official NEDA account (red line), but by collaborators and audience retweets (blue line). The largest number of tweets come from \texttt{@NEDAstaff}, and two private accounts, 
with each tweeting (or retweeting) around 200-300 tweets. Beside the official \texttt{@NEDAstaff} account, the others do not have an official NEDA affiliation. 
In general terms, they are activists who relate to the content of other sources besides NEDA. They often interact with other institutional accounts such as ``Eating Disorder Hope'', ``ADAA'' (Anxiety and Depression Association of America), or ``Mental Health America''.






Finally, in order to compare the users who have engaged with NEDA content, we consider a ``baseline'' set of users who do not necessarily engage with NEDA. We collect a sample of users who have tweeted on February 25 - March 3 on any of diet and health related words. Similarly to the NEDA dataset, we collect the historical tweets of the captured users, resulting in 539,844 tweets of 1,668 different users.

\section{Results}



\subsection{Reach}

Although \texttt{@NEDAstaff} account had 37,567 followers during the NEDAWareness month, their content was retweeted by several popular accounts, dramatically expanding its potential reach. Table \ref{tab:usersfollowers} shows the 20 accounts with the largest number of followers who have retweeted NEDA content. Note that many are accounts of media companies (\texttt{@instagram}, \texttt{@MTV}, \texttt{@MTVNEWS}, \texttt{@Pinterest}), others are magazines (\texttt{@WomensHealthMag}, \texttt{@MensHealthMag}, \texttt{@TeenVogue}), with additional engagement from governmental institutions including The National Institute of Mental Health (\texttt{@NIMHgov}) and U.S. Department of Health \& Human Services (\texttt{@HHSGov}), as well as the Human Rights Campaign (\texttt{@HRC}).

\begin{table}[h]
\centering
\caption{Accounts retweeting NEDAwareness content, ranked by number of followers (in thousands, K).}
\label{tab:usersfollowers}
{
\setlength{\tabcolsep}{12pt}
\begin{tabular}{lr|lr}
\toprule
\textbf{username} & \textbf{\# fol.'s} & \textbf{username} & \textbf{\# fol.'s}\\
\midrule
instagram & 36,665 K & harpersbazaarus & 1,677 K \\
MTV & 15,499 K & seventeen & 1,359 K \\
MTVNEWS & 5,160 K & NIMHgov & 1,153 K \\
WomensHealthMag & 4,581 K & womenshealth & 936 K \\
MensHealthMag & 4,516 K & HRC & 811 K \\

TeenVogue & 3,340 K & HHSGov & 754 K \\
inquirerdotnet & 2,792 K & dosomething & 750 K \\
Ginger\_Zee & 2,340 K & ABC7NY & 653 K \\
Pinterest & 2,337 K & teddyboylocsin & 646 K \\
Jimparedes & 1,751 K & Allure\_magazine & 576 K \\


\bottomrule
\end{tabular}
}
\end{table}

\begin{figure}[h]
    \centering
    \includegraphics[width=0.9\linewidth]{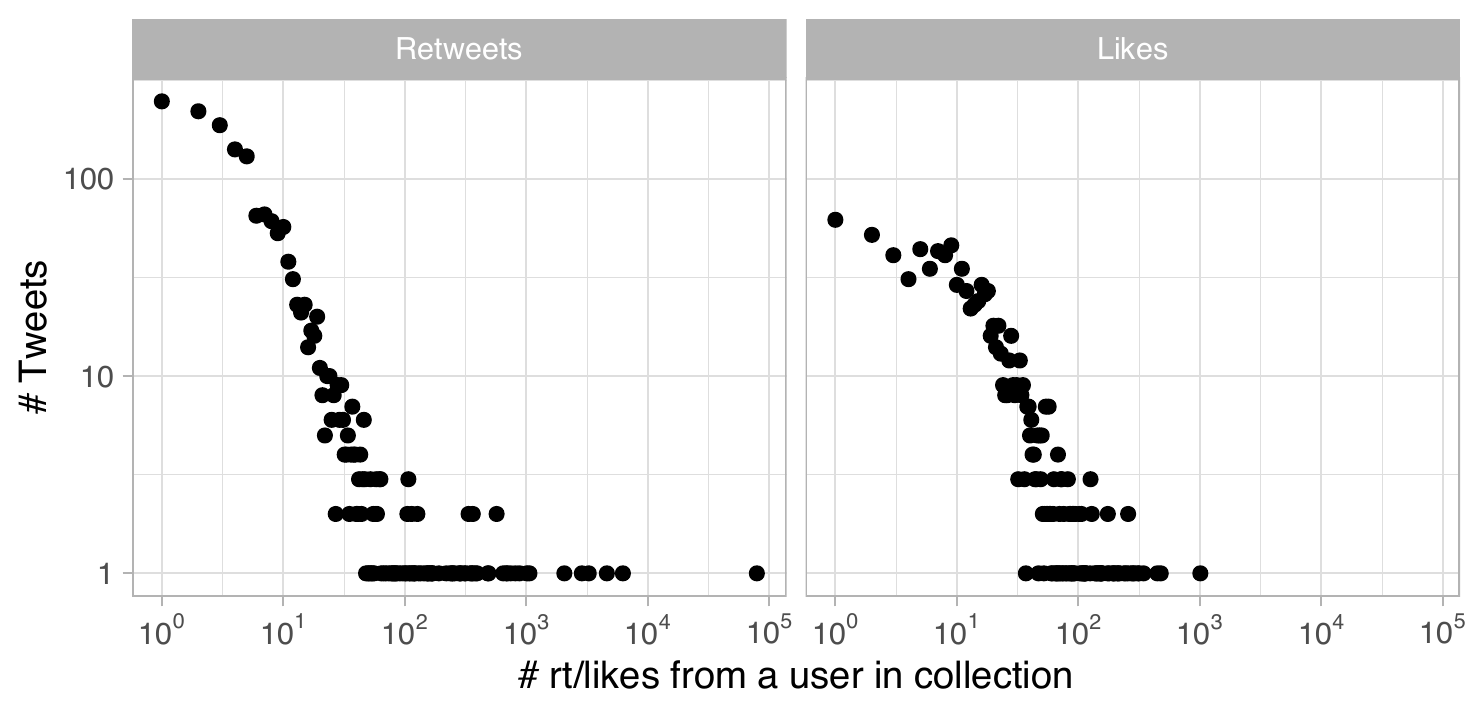} 
    \caption{Distribution of tweets having certain number of retweets (left) and likes (right), log scale.}
    \label{fig:loglogretweetslikes}
\end{figure}

\begin{figure}[t]
    \centering
    \includegraphics[width=0.8\linewidth]{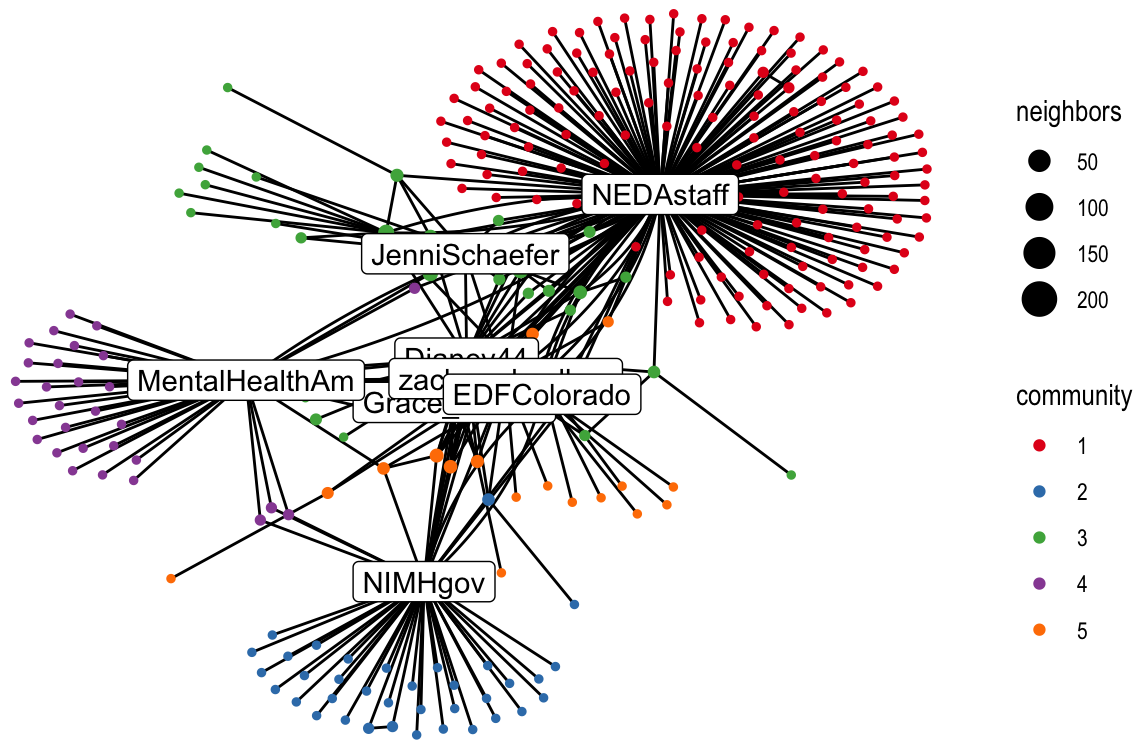}
    \caption{Retweet network GCC with communities.}
    \label{fig:retweetnetwork}
\end{figure}

Besides potential views, we examine the retweet and like statistics of the content in Figures \ref{fig:loglogretweetslikes}. Both are heavy tailed distributions, having a median of 2 likes and 4 retweets, with most posts getting little interaction. 
Furthermore, the Giant Connected Component of the retweet network, in which nodes are users and edges the retweet relationship, can be seen in Figure \ref{fig:retweetnetwork}, colored with five communities identified via the Walktrap algorithm \cite{igraph,ggraph}. NEDA account is at the center of the largest community, followed by \texttt{@NIMHgov} and \texttt{@MentalHealthAm} (Mental Health America, a nonprofit organization). Note that the best reach in terms of retweets was achieved via governmental and nonprofit accounts, despite the more popular media accounts being involved in the conversation, putting in question whether such influencers result in wider reach.

\subsection{Impact}

\subsubsection{Timeline partitioning.}

The continuous nature of Twitter allows us not only to see the unfolding of the NEDAwareness campaign, but also to capture the posting behavior of users involved, both before and after. Here, we ask, to what extent has the language of Twitter users changes after interacting with NEDAwareness content? We take an exploratory approach wherein we examine the posting behavior 15 days before and 15 days after such interaction, as defined below.


Following previous work on health behavior change \cite{west2013cookies}, we define ``day 0'' as the first time a user engages with NEDA content by retweeting it or posting a related tweet. 
Note that we cannot track users who merely saw the NEDAwareness content, as this information is not available through Twitter API. Then we select all users who have tweets published in the 15 days before and after interaction with the NEDA content. We call the 1,746 users who have at least 3 tweets in the 15 days before and 3 tweets in the 15 days after ``day 0'' as ``target users''. For the 2,991 ``baseline users'', we define the 0 day as the first day of NEDAwareness campaign, loosely coupling the time span with that of target users.

\subsubsection{Gender.}

To distinguish impact by gender of the users, we compile a (human) name dictionary with associated genders by combining names extracted from a large collection of Google$+$ accounts \cite{magno2014international} with baby names published by National Records of Scotland\footnote{\url{https://www.nrscotland.gov.uk/statistics-and-data/statistics/statistics-by-theme/vital-events/names/babies-first-names/full-lists-of-babies-first-names-2010-to-2014}} and United States National Security\footnote{\url{https://www.ssa.gov/oact/babynames/limits.html}}, resulting in a dictionary containing 106,683 names. We use this name list to match to user names, as well as apply heuristics (such as having ``Mrs." or ``Mr."). Out of those selected for timeline analysis, 762 users were detected as female, 313 as male, and the remaining 671 as unknown. The baseline users were 855 female, 748 male and 1388 unknown, having a better coverage of the male gender.

\subsubsection{Text modeling.}

Instead of considering all words in this content, we group them using Linguistic Inquiry and Word Count (LIWC) dictionary \cite{pennebaker2001linguistic}, which has 72 categories of words grouped into (1) standard linguistic process, (2) psychological process, (3) relativity, and (4) personal concerns. 
We exclude categories dealing with basic grammar (parts of speech) and high-level summary ones for which more focused ones were available. Thus, for the present study, we select 51 categories including self-references (I, we, you, shehe), emotion (posemo, negemo, anxiety, anger), health and body (feel, body, health, sexual), psychology (focus present, focus future, swear) and other life aspects (work, leisure, home, money). Following this categorization, each tweet is represented as a 51-dimensional vector.

\subsubsection{Effect estimation.}

\begin{figure}[t]
    \centering
    \includegraphics[width=0.8\linewidth]{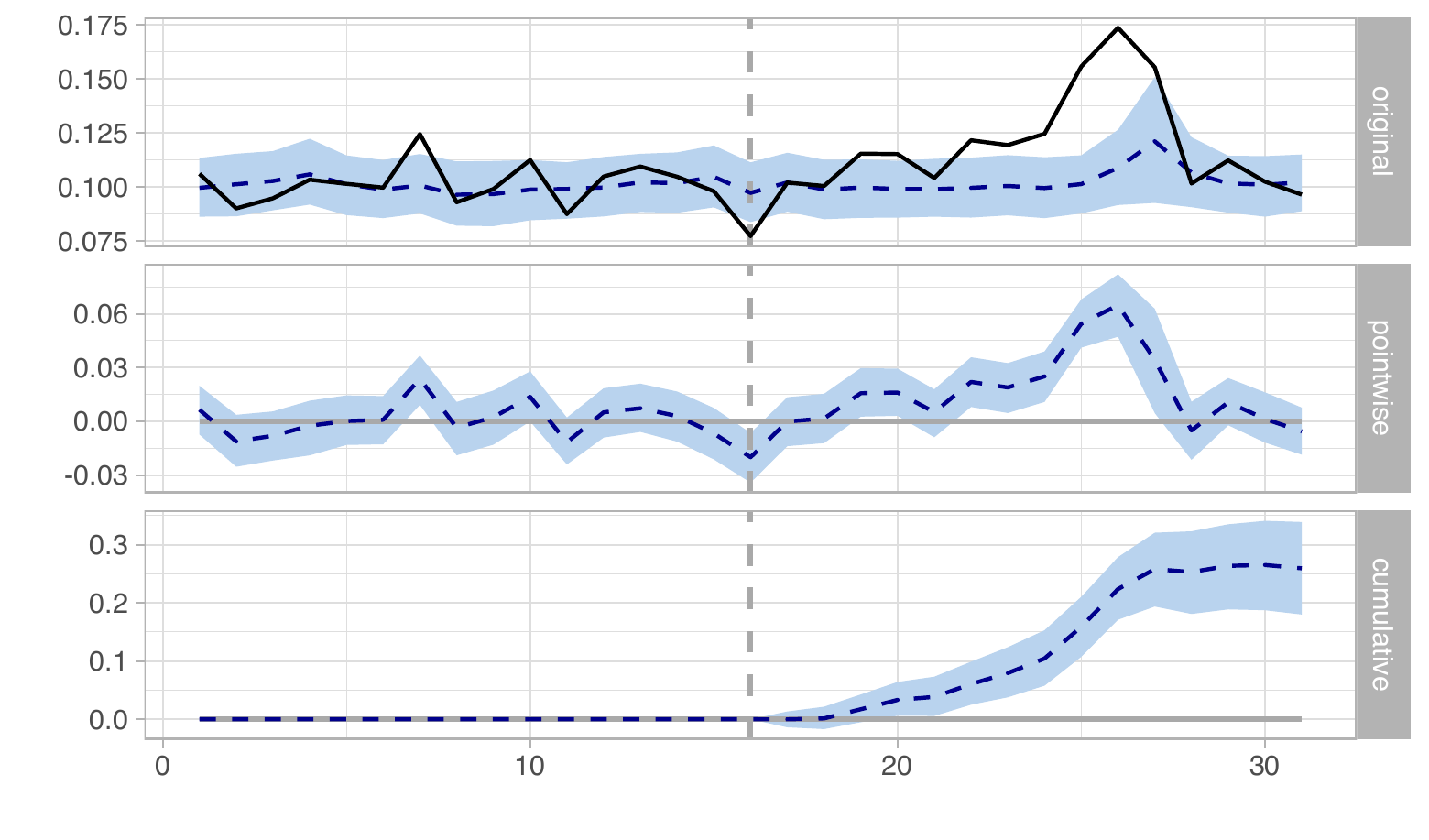}
    \caption{Time series in causal impact analysis for Female category, top: observed tweet rate (solid) and baseline (dashed), middle: difference between the two, bottom: cumulative effect after intervention.}
    \label{fig:femalegraph}
\end{figure}

Using this vector, we measure whether a user's language changes from before to after interacting with NEDAwareness content. We use the Causal Impact Package\footnote{\url{https://google.github.io/CausalImpact/CausalImpact.html}} which estimates the causal effect of some intervention on a behavior over time. This method compares the changes between a response time series (our target users) and a set of control time series (baseline). Given these two series, the package constructs a Bayesian structural time-series model that builds a prediction of the time series if the intervention had never occurred, and compares it to the actual outcome \cite{causalimpact}. For instance, the first panel Figure \ref{fig:femalegraph} shows the actual tweet rate for the Female LIWC category as a solid line and the baseline tweet rate as a dashed line. The second shows the difference between observed data and baseline, and the third shows the cumulative effect of the intervention. In this example, we can observe that the rate of tweets having Female LIWC category is higher than that for the baseline.

Table \ref{tab:impacteffect} shows the categories which have the magnitude of relative effect of 1\% or more and $p$-value $<$ 0.05 in the overall dataset. For these categories, users who have interacted with the NEDA content have changed the way they tweeted after the intervention beyond the changes in the general trend. We also show the effects for each gender separately -- we observe that the effect is not evenly distributed between the genders. 

\begin{table}[h]
\centering
\caption{Relative effect of interaction with NEDA content upon users' use of LIWC categories (with example words in parentheses). Significance: . $p<0.05$, * $p<0.01$.}
\label{tab:impacteffect}
{
\begin{tabular}{l@{\hskip 0.5in}rl@{\hskip 0.2in}rl@{\hskip 0.2in}rl@{\hskip 0.2in}rl}
\toprule
\textbf{Word Category} & \multicolumn{2}{c}{\textbf{All}} & \multicolumn{2}{c}{\textbf{Female}} & \multicolumn{2}{c}{\textbf{Male}} & \multicolumn{2}{c}{\textbf{Unkn.}} \\
\midrule

Female (women, her, she) & 17.4 & * & 13.1 & * & 24.3 & * & 24.0 & * \\
Anxiety (risk, stress, upset) & 7.6 & * & 13.0 & * & -2.7 &  & 10.9 & * \\
Family (family, daughter, families) & 6.5 & * & 7.0 & * & 4.2 &  & 15.2 & * \\
Money (donate, donation) & 6.0 & * & 4.9 & * & 14.0 & * & 4.1 &  \\
Religion (church, goddess) & 5.2 & . & 0.7 &  & 8.3 &  & 12.5 & * \\
Achievement (team, queen, celebrat*)  & 3.8 & . & 5.2 & . & -2.9 &  & 4.0 & . \\
They (they) & 3.4 & . & 5.8 & * & 1.6 &  & 0.7 &  \\
Negate (don't) & 2.9 & * & 4.4 & * & -6.6 & . & 6.5 & . \\
Health (maternity) & 2.5 & * & 4.2 & . & 9.0 & * & -2.5 &  \\
Power (help, threat, terror) & 2.5 & . & 1.0 &  & 1.3 &  & 4.5 & * \\
Negative emotions (risk, stress, upset) & 2.1 & . & 3.9 & * & -3.4 &  & 0.6 &  \\
Informal (retweet, twitter, fb) & 1.1 & . & 0.0 &  & 3.2 & * & 0.7 &  \\
See (look, bright, show) & -2.0 & . & 2.0 &  & 0.2 &  & -7.9 & * \\
Ipron (I) & -1.5 & . & -0.2 &  & -3.0 & . & -1.2 &  \\
Discrepancy (inadequa*) & -2.0 & . & -0.6 &  & -9.5 & * & 1.2 &  \\
You (You) & -2.2 & . & -3.1 & . & 1.0 &  & -2.9 & . \\
Different (different, didn't) & -2.7 & * & -0.5 &  & -5.6 & . & -0.8 &  \\
Positive emotion (share, sharing, help) & -3.3 & * & -3.1 & * & -7.5 & * & -1.4 &  \\
Tentative (unsure, confusing, confused) & -3.3 & * & -0.9 &  & -3.1 &  & -6.0 & * \\
She/he (his, he, her, she) & -7.0 & . & -7.9 & . & -1.8 &  & -4.5 &  \\
Affiliation (we, our, us) & -7.2 & * & -5.9 & * & -8.9 & * & -6.9 & * \\

\bottomrule
\end{tabular}
}
\end{table}

\vspace{-0.5cm}
\subsubsection{Changes in language.}

As can be seen from the table, the category showing most change after the intervention is \emph{Female}, containing words such as \emph{women}, \emph{she}, \emph{her}, etc. For example, the following tweet talks about the trans woman identity and emphases the word \emph{women}: ``\emph{rt (USER): trans women are women. trans women are women. trans women are women. trans women are women. trans women are women. trans women are...}''. In particular, the time series for \emph{Female} category showed an increase of +17\% (95\% interval [+11\%, +23\%]). This means that the positive effect observed during the intervention period is statistically significant and unlikely to be due to random fluctuations. Interestingly, this effect is strong across users of all genders, including unknown. Some of the increase in this category can be attributed to the International Women's Day that happens on March 8\footnote{\url{https://en.wikipedia.org/wiki/International\_Women's\_Day}}. Note that although this topic is not directly connected to eating disorders, users who have interacted with NEDA content are more likely to tweet about this holiday than the control group, indicating a heightened awareness of the holiday, and possible women's rights issues associated with it.

\begin{figure}[t]
    \centering
    \subfloat[Family: Female]{\includegraphics[width=0.24\linewidth]{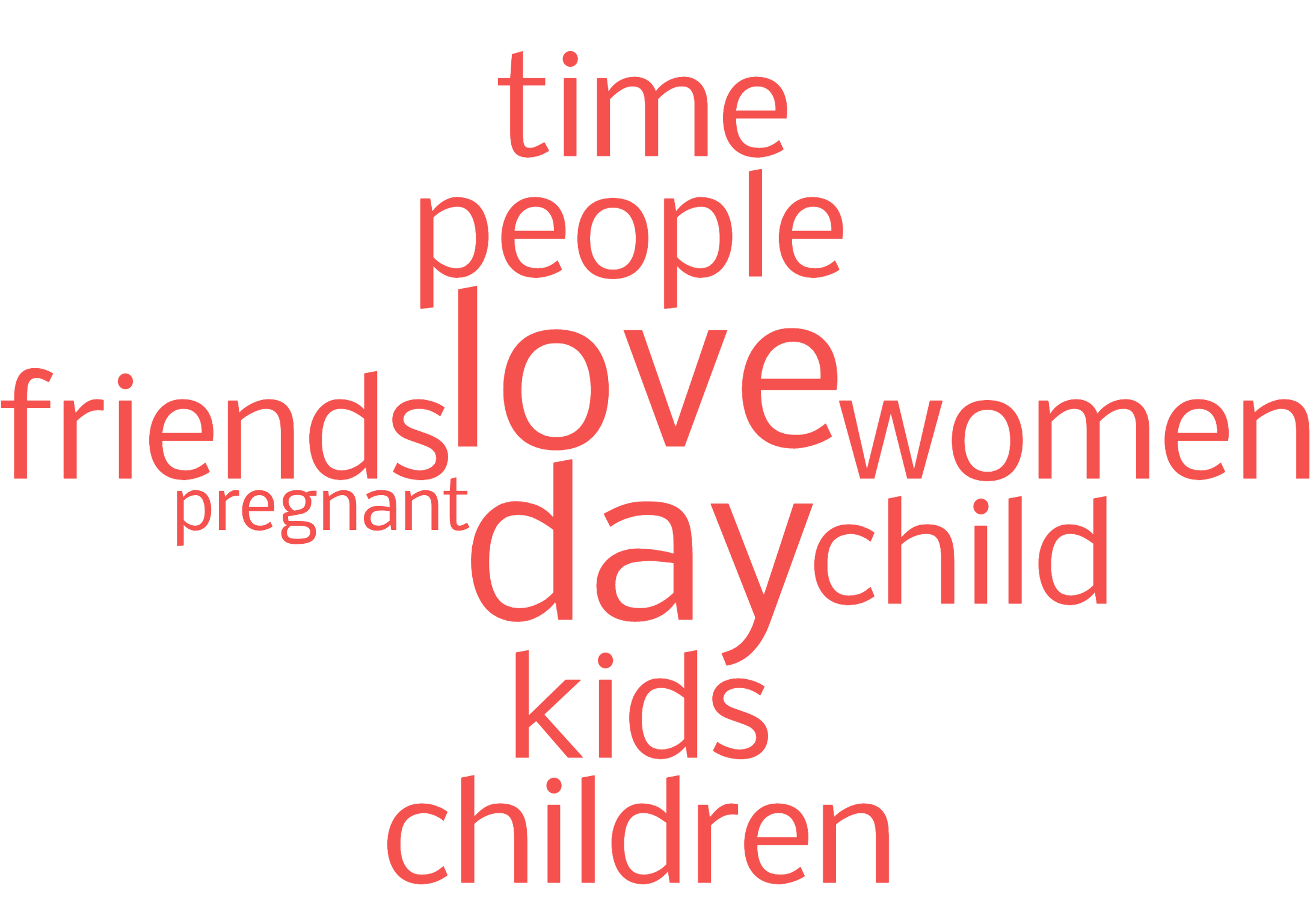}}\hspace{0.2cm}
    \subfloat[Family: Male]{\includegraphics[width=0.23\linewidth]{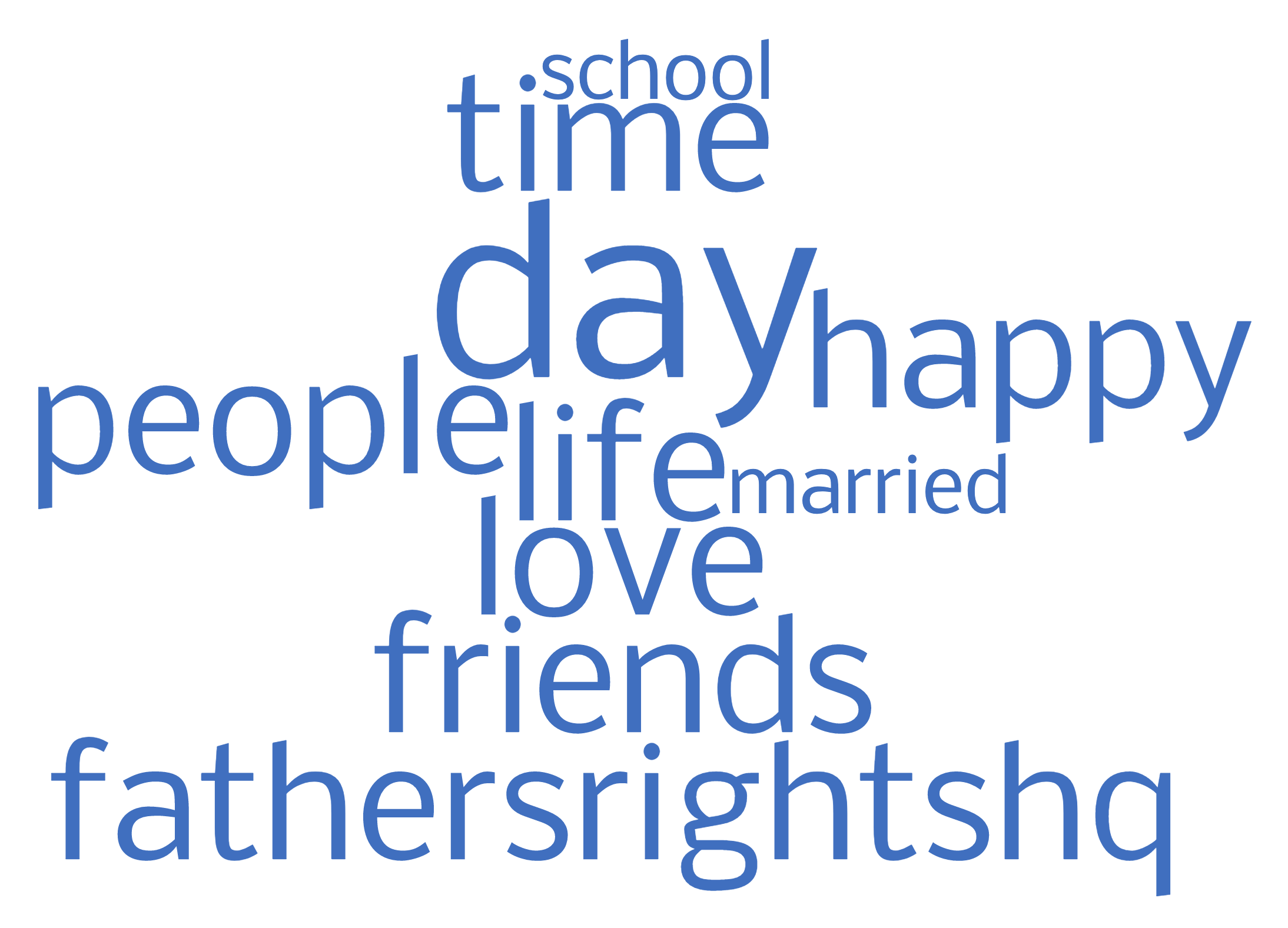}}\hspace{0.4cm}
    \subfloat[Anxiety: Fem.]{\includegraphics[width=0.21\linewidth]{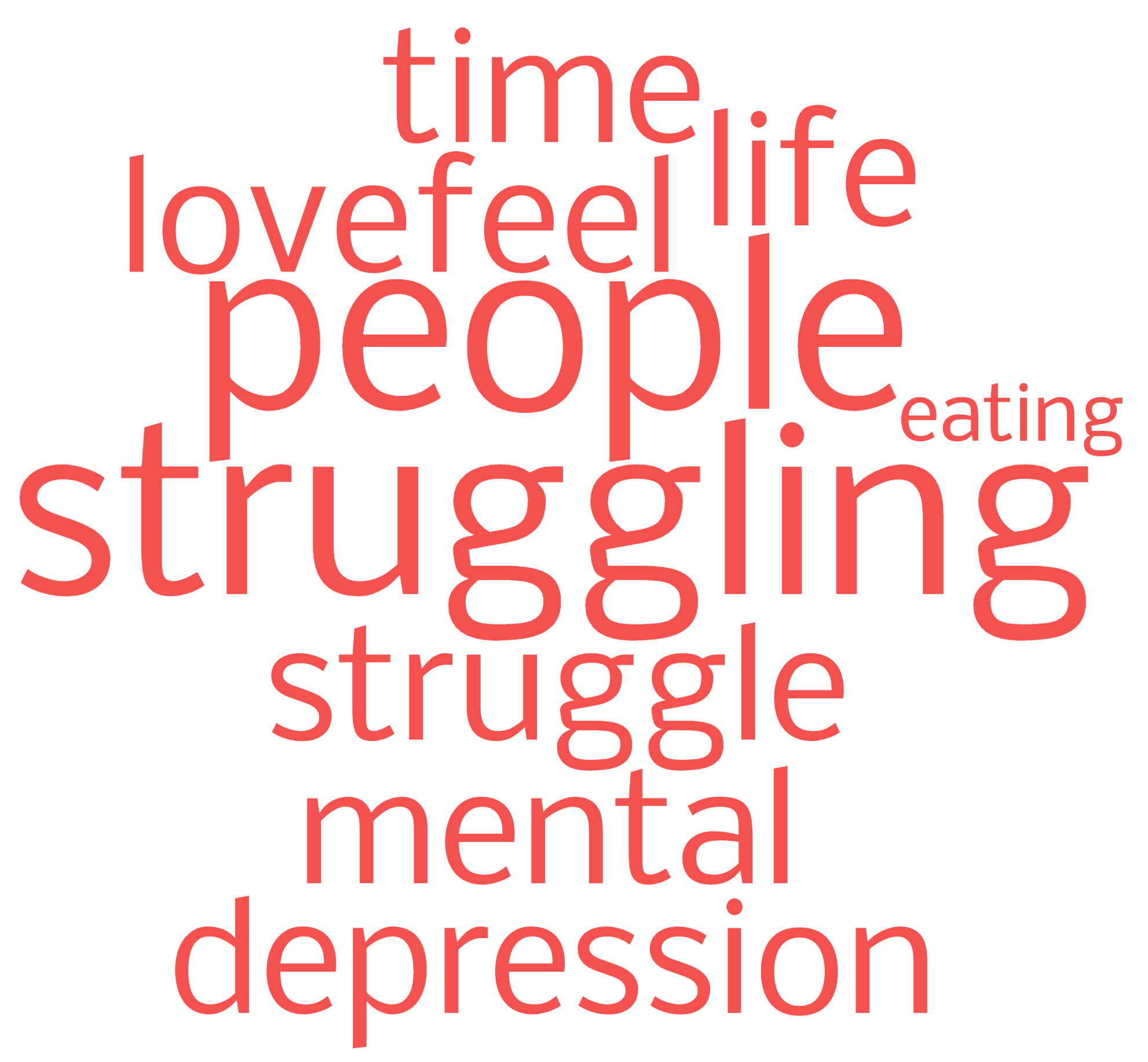}}\hspace{0.2cm}
    \subfloat[Anxiety: Male]{\includegraphics[width=0.22\linewidth]{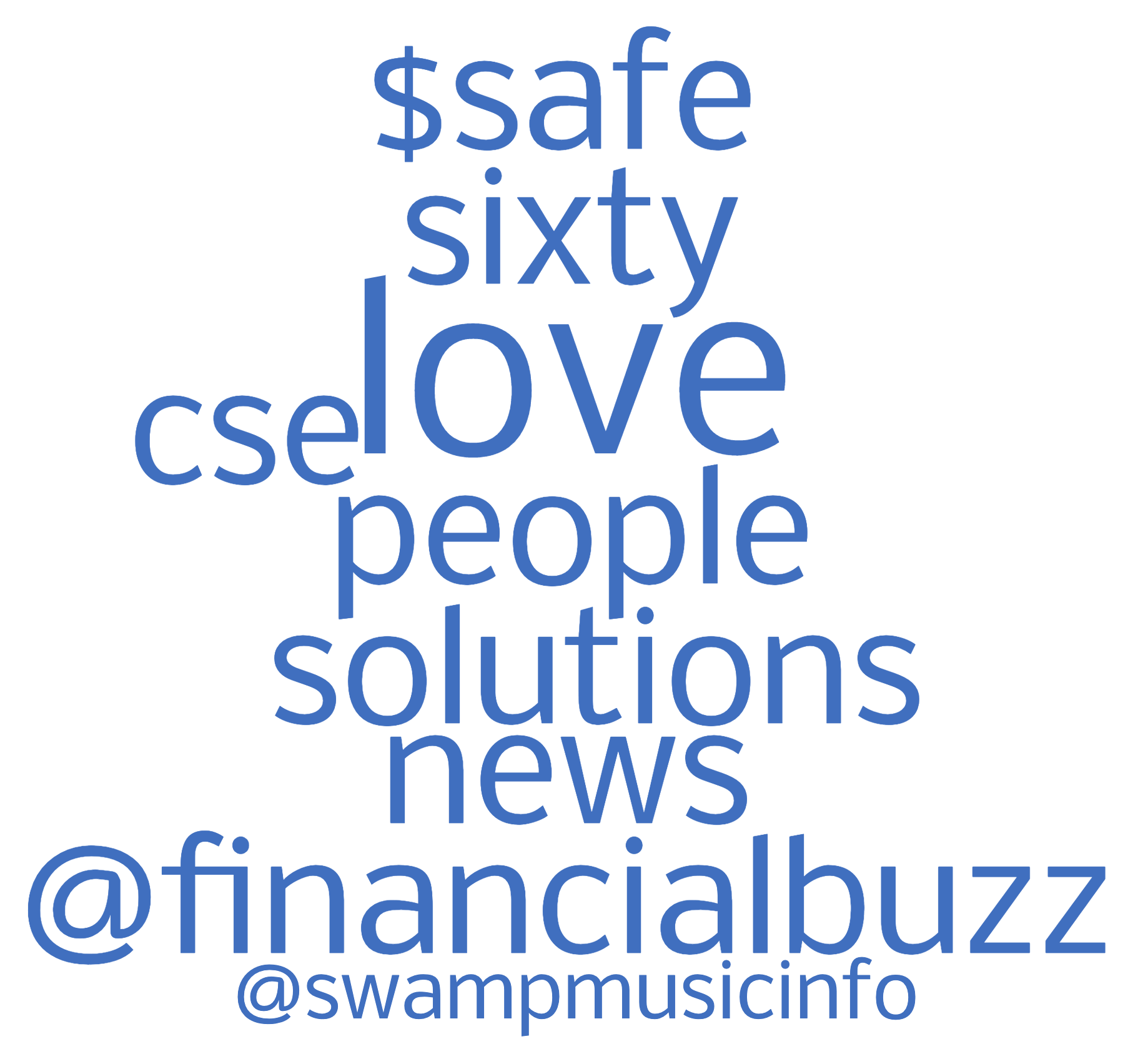}}
    \caption{Top 10 words associated with family and anxiety categories, by gender.}
    \label{fig:wordclouds}
    \vspace{-.5cm}
\end{figure}

Second most affected category is \emph{Anxiety} category, with an increase of +8\% ([+3\%, +12\%]). The words most used in this category are \emph{risk}, \emph{stress}, \emph{upset} and \emph{worry}. Interestingly, the words less used were \emph{confusing}, \emph{horrible} and \emph{doubts}. For instance, users share their feelings, such as in this example: ``\emph{currently I am restless, scared, mistrustful, rattled, insecure, frightened, impatient, anxious.}''. The category is significantly different for female users (as well as unknown gender), and not male. Figure \ref{fig:wordclouds}(c,d) shows the top 10 words associated with anxiety category words for female and male users. We find female users mentioning words ``struggling'' and ``struggle'', as well as ``mental'', ``depression'' and ``eating''. On the other hand, many keywords on male side are associated with finances, such as ``@financialbuzz'' and ``\$safe'' (by convention, ``\$'' precedes tickers or financial information), as well as ``cse'' (National Child Sexual Exploitation Awareness).

Third most affected category is \emph{Family}, with an increase of +6\% ([+2\%, +11\%]). This category includes words related to the family. Interestingly, the words within that category which are most used are again closely related to women, which are \emph{ma}, \emph{daughter}, and \emph{family}. Note that \emph{wife}, \emph{bro} and \emph{daddy} are the least used within this category. For example, the following tweet emphasizes the female members of families and their needs: ``\emph{rt (user): they are our sisters, nieces, cousins, daughters, aunts, granddaughters, wives, mothers, grandmothers, friends... they need...}''. Figure \ref{fig:wordclouds}(a,b) shows the top 10 words most associated with the family category (those found in the context of tweets also containing words from family category) for the male and female gender. We find that, whereas women mention ``kids'' and ``pregnant'', for men the emphasis is more on ``friends'', ``life'', and ``time''. We also notice a keyword related to ``FathersRightsHQ'', which posts a mixture of political news and mental issues associated with family.

On the bottom of Table \ref{tab:impacteffect}, we find categories which are used less by the users after interacting with NEDA content. It is more difficult to provide concrete examples of content \emph{not} posted, but we draw the reader's attention to the fact that these categories are \emph{Affiliation} (ex: \emph{boyfriend}, \emph{our}, \emph{together}), \emph{She/he} (ex: \emph{he}, \emph{she}, \emph{herself}), and \emph{Positive emotion} (ex: \emph{amazing}, \emph{favorite}, \emph{sharing}).

\begin{figure}[t]
    \centering
    \includegraphics[width=0.6\linewidth]{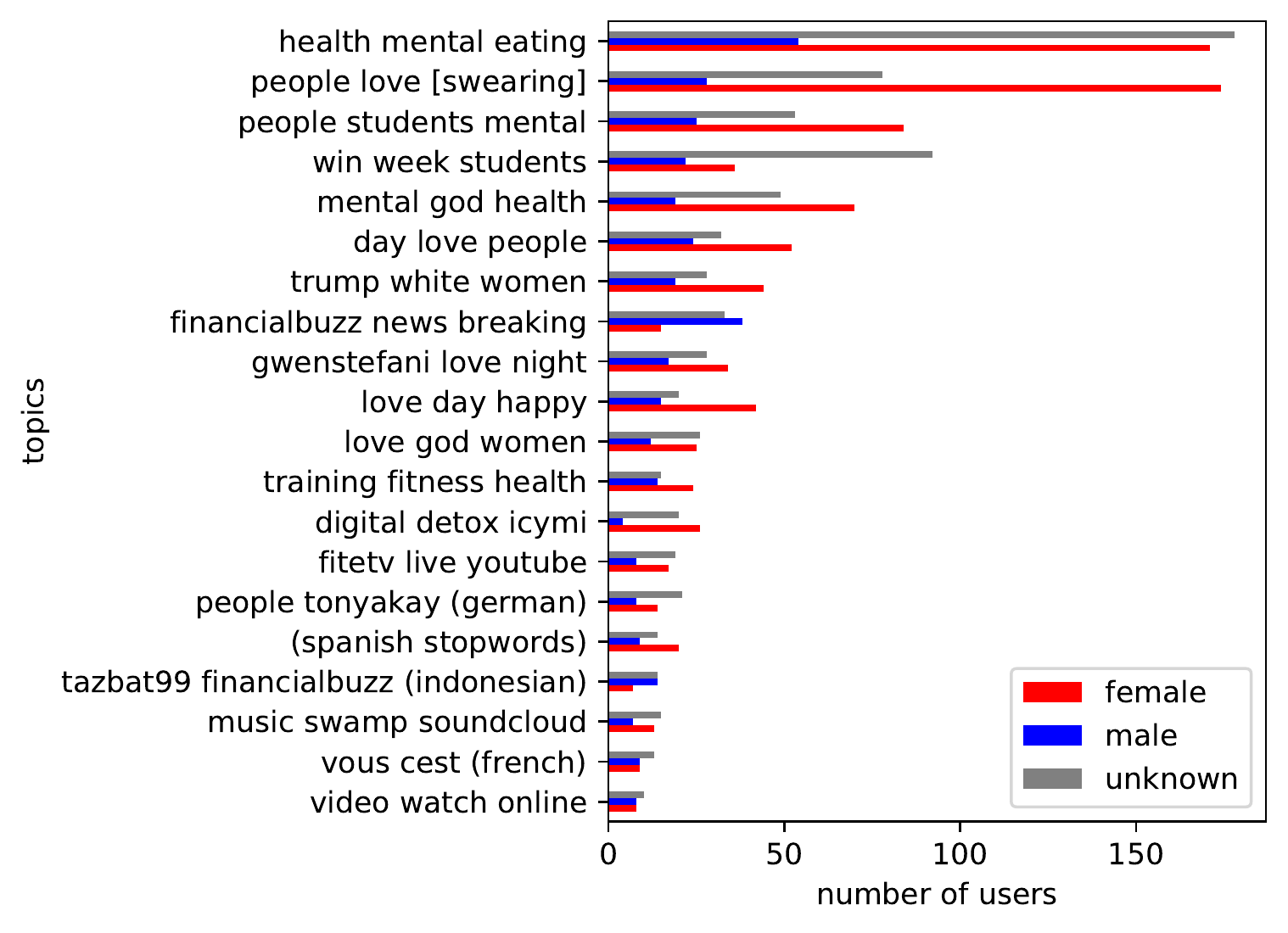}
    \caption{Number of users assigned to LDA topics, per gender.}
    \label{fig:ldatopicspergender}
\end{figure}

Finally, we model the text produced by all of these users in the 15 days after NEDAwareness event via LDA at k=15 topics (selected manually for greatest cohesion). Figure \ref{fig:ldatopicspergender} shows the prevalence of each topic in each detected gender, with each topic signified by the top descriptive terms. Topics are assigned by the greatest likelihood to the aggregated tweets for each user. As there are more female users, we find most topics to be dominated by them, except the topics around ``financialbuzz''. Other topics having a larger proportion of males are around ``students'', ``training'' and ``fitness'', as well as around music (``gwenstefani''). As we can see, the main topics of family, anxiety, and womanhood do not show up in these topics, illustrating the need for finer-grained analysis above.

\section{Discussion \& Conclusions}

Modeling user behavior through online self-expression is an important complement to the traditional survey-based methods of behavior change evaluation, extending the reach of analysis at a low cost. Social informatics community has recently focused on health recommender systems and gamification, for instance, estimating the factors determining a mobile user's perception of the recommendation \cite{torkamaan2019rating}, testing gamified persuasive messaging for behavior change \cite{ciocarlan2018kindness}, and developing game design guidelines for improving subjective wellbeing \cite{coicarlan2017qualitative}. This work extends the purview of the intervention to social media -- a platform increasingly used for health messaging -- and proposes an unobtrusive methodology for tracking change in self-expression, as compared to pre-intervention levels, as well as in comparison to a control group.

Quantitative analysis of the content's reach has shown that, despite accounts with large followings being involved in the campaign's promotion, the most engagement in terms of retweets have come from government and nonprofit organizations, (National Institute of Mental Health) and nonprofit (Mental Health America) putting in question the effectiveness of influencers for the promotion of health messaging (the effectiveness of such influencers may be further studied through the lens of social-psychological theory \cite{mcneill2014understanding}). 

Further, our analysis of the posts by the users who have interacted with the campaign's messaging has revealed several important trends: 

\begin{itemize}

\item After the campaign, these users began speaking more about \emph{women} and \emph{family} (latter also more inclined to female members), indicating a general concern over womanhood in the context of mental health, despite the focus of the campaign on diversity. This finding supports latest research, which finds that although disordered eating affects both genders, women are more likely to report binge eating and fasting \cite{striegel2009gender}. Further, keywords associated with children and childbearing was expressed much more by users identified as female, suggesting the continuation of the concern for child-rearing to be largely the purview of women. However, we do note the popularity of content associated with ``fathers rights'' in men's tweets (note children are not in the list of top family terms for male accounts).

\item Secondly, despite positive body acceptance messaging of the campaign, we find a marked increase in anxiety-related and decrease in positive emotion words, suggesting the audience of the intervention has the need to share negative experiences. Indeed, mental health self-disclosure has been observed on several social media platforms, and been compared to a virtual ``support group'' \cite{naslund2016future}. How large organizations like NEDA fit into such a community is an interesting research question. In particular, the rise in anxiety-related words was more pronounced for users identified as female, instead of male ones (although the smaller number of male users may have played a role in statistical significance calculations). The outright mentions of ``depression'' and ``mental'' keywords by female users may point to \emph{psychological openness} that has been recorded in women to seek help from mental health professionals \cite{mackenzie2006age}.

\item This leads to our third observation: the decrease in the use of words such as \emph{boyfriend}, \emph{our}, and \emph{together} and others in the \emph{affiliation} category, indicating a comparative lack of social engagement, as expressed by the users. Unlike in other word categories discussed above, this change is more prevalent for male users. As social isolation has been shown to affect vulnerable youths \cite{storch2007peer}, it may be an important component of well-being to track.

\end{itemize}

A number of notable limitations must be mentioned. First, all purely observational studies are limited to public behaviors people choose to share with others. To complement the public view of an individual with the private, we plan on extending this study with surveying, diaries, and other traditional techniques (such as in \cite{attai2015twitter} which evaluated the impact of a breast cancer education drive on Twitter). Second, the time span of the analysis should be lengthened beyond 15 days studied here to measure the long-term behavior of the subjects, as well as tracking their ``information diet'' \cite{kulshrestha2015characterizing} that may reinforce or undermine the desired behavior. Third, the demographics of the impacted population are unclear, with limited location and gender information available on Twitter, and further studies in technology usage will allow for a more precise estimates of message exposure \cite{mccloud2017cancer}. Finally, this work does not utilize the images posted by the users, and in future work we plan to extract further features from multimedia, following recent work in \cite{guntuku2019twitter}.

Importantly, mental health research deals with potentially vulnerable populations, and whereas in this work only the largest accounts were revealed and example tweets rephrased as much as possible for de-identification, privacy is an ongoing concern. To limit the exposure of the individuals involved, the data will be made available to other researchers in anonymized fashion, and in accordance with EU General Data Protection Regulation (GDPR).


\balance

\bibliographystyle{splncs04}
\bibliography{neda}

\end{document}